\documentstyle[12pt]{article}
\begin{document}
\def\baselinestretch{1.3}
\topmargin=0in                          
\headheight=0in                         
\headsep=0in                    
\textheight=220mm                               
\footheight=3ex                         
\footnotesep=.5cm
\footskip=4ex           
\textwidth=150mm                                
\hsize=150mm                            
\parskip=\medskipamount                 
\parindent=1cm
\lineskip=12pt                          
\def\bra#1{\langle #1 |}
\def\ket#1{| #1\rangle}                 
\def\7#1#2{\mathop{\null#2}\limits^{#1}}        
\def\ast{\displaystyle *}
\let\ve=\varepsilon
\def\beee{\begin{equation}}
\def\eeee{\end{equation}}  
\def\dggg{^{\dagger}}
\def\qnu{[n_u]_q}
\def\qnv{[n_v]_q}
\def\nqq{[n]_q}
\def\nqp{[n-1]_q}
\oddsidemargin=.0in
\evensidemargin=.0in
\thispagestyle{empty}
\bibliographystyle{unsrt} 
\begin{center}
{\LARGE\bf Theories of Violation of Statistics}\\
\vspace{5mm}
O.W. Greenberg\footnote{email address, owgreen@physics.umd.edu}\\
{\it Center for Theoretical Physics\\
Department of Physics \\
University of Maryland\\
College Park, MD~~20742-4111}\\
\vspace{10mm}
{\normalsize\it Dedicated to George Abraham Snow, 
August 24, 1926-June 24, 2000.}\\
\vspace{10mm}
University of Maryland Preprint PP-00-085\\
Talk given at the International Conference on the Spin-Statistics Connection\\
and Commutation Relations:  Experimental Tests and Theoretical Implications,\\
Capri, Italy, May 31-June 4, 2000.\\
To be published by the American Institute of Physics\\
in the Proceedings of the Conference,
ed. R.C. Hilborn and G. Tino.
\end{center}

\begin{abstract}      
I discuss theories of violations of statistics,
including intermediate statistics, parastatistics, parons,
and quons.  I emphasize quons, which allow
small violations of statistics.  I analyze the quon
algebra and its representations, implications of the
algebra including the observables allowed by the superselection rule
separating inequivalent representations of the symmetric group,
the conservation of statistics rules, and the rule for composite
systems of quons.  I conclude by raising the question of possible origins
of violations of statistics and of the level at which violations should
be expected if they exist.
\end{abstract}

\section{Introduction}

As far as I know this is the first international conference devoted entirely
to the relation of spin and statistics and to the investigation of possible
small violations of statistics.  My purpose in this talk is to give an overview
of theoretical issues connected with violations of statistics.  I have divided
the talk into five parts:  (1) general theoretical remarks, (2) types of experiments
to detect violations, (3) attempts to violate statistics, (4) quons, the best
formalism so far to describle small violations, and (5) summary and open questions.

\section{General theoretical remarks}

The general principles of quantum theory do not require that all particles
be either bosons or fermions.  This restriction requires an additional postulate
which A.M.L. Messiah\cite{mes} named the ``symmetrization postulate,'' which I quote as
Messiah defined it: ``The states of a system containing $N$ identical particles
are necessaribly either all symmetrical or all antisymmetrical with respect to
permutations of the $N$ particles.''  The symmetrization postulate can be restated
as: ``All states of identical particles are in one-dimensional representations
of the symmetric group.''  With Messiah's definition the spin-statistics
connection, that integer spin particles are bosons and odd-half-integer spin
particles are fermions, is a separate statement.

Messiah and I gave a detailed discussion of quantum mechanics without the 
symmetrization postulate\cite{mesowg}.  We emphasized that without the 
symmetrization postulate, a set of one-body measurements is never a maximal
set; one needs additional measurements to fix the state of the system.  Further
there is a superselection rule separating states in inequivalent representations
of the symmetric group.  If identical particles can occur in states that violate
the spin-statistics connection their transitions must occur in the same 
representation of the symmetric group.  For example, in the experiment of 
Deilamian, et al\cite{dei} which looked for anomalous helium atoms in which
the two electrons violated the exclusion principle and were in the symmetric 
state, the search was for transitions among the symmetric states rather than
between symmetric and antisymmetric states.  This point was also made by R.
Amado and H. Primakoff\cite{ama}.  In addition if one assumes that charged particles
couple universally to the electromagnetic field, then the transitions among the anomalous
states occur at the normal rate, so that isolated atoms will be in the lowest state of
the anomalous system.  Since the symmetrization postulate is not an intrinsic part
of quantum theory, this postulate must be subjected both to theoretical study and to 
experimental tests.  Quantitative tests require a theory in which the symmetrization
postulate does not have to hold and in which the violation of the postulate is reflected
in a parameter that departs from its standard value at which 
the symmetrization postulate and
the spin-statistics connection do hold.  It is certainly possible that violations of
statistics are extremely small and require high-precision tests to be observed.  This
conference brings together leading workers in this search. 

Because the notion that particles are identical requires that the Hamiltonian
and, indeed, all observables must be symmetric in the dynamical variables associated
with the identical particles, the observables can't change the permutation symmetry
type of the wave function.  In particular one can't introduce a small violation of
statistics by assuming the Hamiltonian is the sum of a statistics-conserving and a
small statistics-violating term,
\beee
H=H_S+\epsilon H_V
\eeee
as one can for violations of parity, charge conjugation, etc.  Violation of statistics
has to be introduced in a more subtle way.

If charged particles couple universally to the electromagnetic field, then
there can't be two kinds of--say--electrons, ``red'' electrons and ``blue''
electrons, because then the lowest order pair production cross section,
\beee
\sigma(\gamma X \longrightarrow e^+ e^- X)
\eeee
would double. A high-precision measurement is not needed to rule this out.

A convenient way to parametrize violations or bounds on violations of 
statistics uses the two-particle density matrix.  For fermions,
\beee
\rho_2=(1-v_F)\rho_a+v_F\rho_s;
\eeee
for bosons,
\beee
\rho_2=(1-v_B)\rho_s+v_B\rho_a;
\eeee
in each case the violation parameter varies between zero if the statistics is
not violated and one if the statistics is completely violated.
\section{Types of experiments}

There are three basic types of experiments to detect violations of statistics:
(1) transitions among anomalous states--these can occur in solids, liquids or gases,
(2) accumulation of particles in anomalous states, and (3) deviations from the usual statistical
properties of the identical particles.  Since, as mentioned earlier, a superselection rule
prevents transitions between normal and anomalous states, experiments searching for such
transitions do not provide a valid test of violation of statistics.  

Transitions among
anomalous states can provide a very sensitive test, since in some cases a single such
transition can be observed.  The prototype of this kind of test is the experiment of Maurice
and Trudy Goldhaber\cite{gol}.  They asked the qualitative question, ``Do the
electrons from nuclear beta decay obey the same exclusion principle as electrons in
atoms?''  They knew that electrons from each source have the same charge, spin, and 
mass, etc., i.e.\ that the single-electron states in each case are identical, but 
there was no evidence that the many-electron states from each source are identical.
They devised the following ingenious test: they let beta decay electrons from a
nuclear source fall on a block of lead.  They argued that if the many-electron states
were not identical then the nuclear beta decay electrons would not obey the same
exclusion principle as the electrons in the lead atoms.  Then the beta decay electrons
would not see the $K$ shells in the lead atoms as filled and could fall into the
$K$ shells and would emit $x$-rays.  A single such $x$-ray could be observed.
They saw no such $x$-rays above background and thus answered their qualitative
question in the affirmative.  I estimate that their experiment gave the bound
$v_F\leq 5 \times 10^{-2}$ for electrons.

E. Ramberg and G.A. Snow\cite{ram} developed this experiment into one which yields a
high-precision bound on violations of the exclusion principle.  Their idea was
to replace the natural $\beta$ source, which provides relatively few electrons,
by an electric current, in which case Avogadro's number is on their side.  The
possible violation of the exclusion principle is that a given collection of
electrons can, with different probabilities, be in different permutation
symmetry states.  The probability to be in the ``normal'' totally antisymmetric
state presumably would be close to one, 
the next largest probability would occur for the
state with its Young tableau having one row with two boxes, etc.  The idea of
the experiment is that each collection of electrons has a possibility of being
in an ``abnormal'' permutation state.  If the density matrix for a conduction 
electron together with the electrons in an atom has a projection onto such an
``abnormal'' state, then the conduction electron will not see the $K$ shell of
that atom as filled. Then a transition into the $K$ shell with $x$-ray emission
is allowed.  Each conduction electron which comes sufficiently close to a given
atom has an independent chance to make such an $x$-ray-emitting transition, and
thus the probability of seeing such an $x$-ray is proportional to the number of
conduction electrons which traverse the sample and the number of atoms which the
electrons visit, as well as the probability that a collection of electrons can
be in the anomalous state.  Ramberg and Snow chose to run 30 amperes
through a thin copper strip for about a month.  They surrounded the experiment
with veto scintillators to remove background $x$-rays.
They estimated the energy of the modified
$x$-rays which would be emitted due to the transition to the $K$ shell.  No 
excess of $x$-rays above background was found in this energy region.  Ramberg
and Snow set the limit 
\beee
v_F \leq 1.7 \times 10^{-26}               
\eeee
for electrons.  This is high precision indeed!

The Ramberg-Snow experiment may seem discouraging for the discovery of generalizations
of bose and fermi statistics; however there are small numbers in physics which, if
necessary, can
occur in degree greater than one.  For example the ratios
\beee
\frac{m_{proton}}{M_{Planck}}\sim 10^{-19}, ~{\rm and} ~
\frac{G_Nm_e^2}{e^2}\sim 10^{-43}
\eeee
can provide numbers smaller than the Ramberg-Snow bound.  In addition new physics
effects such as violations of Lorentz invariance, spacetime discreteness, 
spacetime noncommutativity, etc.\ may provide small effects.
Mohapatra and I gave an early survey of experimental bounds on violations of 
statistics\cite{owg-rnm}.

Composite structure can mimic violations of statistics.  This is not what I am
considering here.

\section{Attempts to violate statistics}

\subsection{Gentile's ``intermediate statistics''}

The first attempt to go beyond bose and fermi statistics seems to have been
made by G. Gentile\cite{gen} who suggested an 
``intermediate statistics'' in which at
most $n$ identical particles could occupy a given quantum state.  In
intermediate 
statistics, fermi statistics is recovered for $n=1$ and bose statistics
is recovered for $n\rightarrow \infty$; thus intermediate statistics 
interpolates between fermi and bose statistics.  However Gentile's
statistics is not a proper quantum statistics, because the condition of having
at most $n$ particles in a given quantum state is not invariant under change
of basis\cite{owgphysica}.  For example, for intermediate statistics with $n=2$, the state
$|\psi \rangle=|k,k,k \rangle$ does not exist; however, the state $|\chi
\rangle= 
\sum_{l_1,l_2,l_3}U_{k,l_1}U_{k,l_2}U_{k,l_3}|l_1,l_2,l_3 \rangle$, obtained
from $|\psi \rangle$ by the unitary change of single-particle basis, 
$|k \rangle ^{\prime}=\sum_{l}U_{k,l}|l \rangle$ does exist.
By contrast, parafermi statistics of order $n$ which I discuss just below is
invariant under change of basis\cite{gre}.  Parafermi statistics of order 
$n$ not only
allows at most $n$ identical particles in the same state, but also allows
at most $n$ identical particles in a symmetric state.  In the example just
described, neither $|\psi \rangle$ nor $|\chi \rangle$ exist for parafermi
statistics of order two.

\subsection{Parastatistics}

H.S. Green\cite{gre} proposed the first proper quantum statistical
generalization of bose and fermi statistics.  Green noticed that the commutator
of the number operator with the annihilation and creation operators is the same
for both bosons and fermions
\beee
[n_k, a\dggg_l]_-=\delta_{kl}a\dggg_l.
\eeee
The number operator can be written
\beee
n_k=(1/2)[a\dggg_k, a_k]_{\pm}+ {\rm const.},
\eeee
where the anticommutator (commutator) is for the bose (fermi) case.  If these
expressions are inserted in the number operator-creation operator commutation
relation, the resulting relation is {\it trilinear} 
in the annihilation and creation operators.  Polarizing the number operator to
get the transition operator $n_{kl}$ which annihilates a free particle in state
$l$ and creates one in state $k$ leads to Green's trilinear commutation relation
for his parabose and parafermi statistics,
\beee
[[a\dggg_k, a_l]_{\pm}, a\dggg_m]_-=2\delta_{lm}a\dggg_k
\eeee
Since these rules are trilinear, the usual vacuum condition,
\beee
a_k|0\rangle=0,
\eeee
does not suffice to allow calculation of matrix elements of the $a$'s and
$a\dggg$'s; a condition on single-particle states must be added,
\beee
a_k a\dggg_l|0\rangle=p \delta_{kl}|0\rangle.
\eeee

Green found an infinite set of solutions of his commutation rules, one for each
positive integer $p$, by giving an ansatz which he expressed in terms of bose and fermi
operators.  Let
\beee
a_k\dggg=\sum_{p=1}^n b_k^{(\alpha) \dagger},~~a_k=\sum_{p=1}^n b_k^{(\alpha)},
\eeee
and let the $b_k^{(\alpha)}$ and $b_k^{(\beta) \dagger}$ 
be bose (fermi) operators
for $\alpha=\beta$ but anticommute (commute) for $\alpha \neq \beta$ for the 
``parabose'' (``parafermi'') cases.  This ansatz clearly satisfies Green's
relation.  The integer $p$ is the order of the parastatistics.  The physical
interpretation of $p$ is that, for parabosons, $p$ is the maximum number of
particles that can occupy an antisymmetric state, while for parafermions, $p$
is the maximum number of particles that can occupy a symmetric state (in
particular, the maximum number which can occupy the same state).  The case $p=1$
corresponds to the usual bose or fermi statistics.
Later, Messiah
and I\cite{owgmes} proved that Green's ansatz gives all Fock-like solutions of
Green's commutation rules.  Local observables have a form analogous to the usual
ones; for example, the local current for a spin-1/2 theory is 
$j_{\mu}=(1/2)[\bar{\psi}(x), \psi(x)]_-$.  From Green's ansatz, it is clear
that the squares of all norms of states are positive, since sums of bose or
fermi operators give positive norms.  Thus parastatistics\cite{del} gives a set of
orthodox theories.  

This is all well and good; however, the violations of statistics provided by
parastatistics are gross.  Parafermi statistics of order two has up to two
particles in each quantum state.  High-precision experiments are not necessary
to rule this out for all particles we think are fermions.

\subsection{The Ignatiev-Kuzmin model}

Interest in possible small violations of the exclusion principle was revived by
a paper of Ignatiev and Kuzmin\cite{ign} in 1987.  They constructed a model of
one oscillator with three possible states: a vacuum state, a one-particle
state and, with small amplitude $\beta$, a two-particle state.  They gave trilinear 
commutation relations for their oscillator. Mohapatra and I noticed that the 
Ignatiev-Kuzmin oscillator could be
represented by a modified form of the order-two Green ansatz.  We suspected that
a field theory generalization of this model having an infinite number of
oscillators would not have local observables and set
about trying to prove this.  To our surprize, we found that we could construct
local observables and gave trilinear relations which guarantee the locality 
of the current\cite{gremoh}.

\subsection{Parons}

Following Ignatiev and Kuzmin we introduced a parameter $\beta$ that gives the
deformation of the Green trilinear commutation relations.  
For $\beta \rightarrow 1$ the relations reduce to those of 
the $p=2$ parafermi field; for $\beta \rightarrow 0$ the double occupancy is
completely suppressed and the theory is equivalent to a fermi theory.  A random
state of two paronic electrons has the violation parameter $\beta^2/2$.
Mohapatra and I checked that the norms are positive for 
states of up to three particles.  At this stage, we were carried away with
enthusiasm, named these particles ``parons'' since their algebra is a
deformation of the parastatistics algebra, and thought we had found a local
theory with small violation of the exclusion principle.  
Unknown to us Govorkov\cite{gov}, using a detailed algebraic argument,
already had shown in
generality that any deformation of the Green commutation relations necessarily
has states with negative squared norms in the Fock-like representation.
For our model, the first such negative-probability state occurs for
four particles in the representation of ${\cal S}_4$ with three boxes in the
first row and one in the second.  We were able to understand Govorkov's result
qualitatively as follows:\cite{gn3} 
Since parastatistics of order $p$ is related by a
Klein transformation to a model with exact $SO(p)$ or $SU(p)$ internal symmetry,
a deformation of parastatistics which interpolates between Fermi and parafermi
statistics of order two would be equivalent to interpolating between the trivial
group whose only element is the identity and a theory with 
$SO(2)$ or $SU(2)$ internal symmetry.  This is impossible, since there is no
such interpolating group.

\subsection{The Doplicher-Haag-Roberts analysis}

S. Doplicher, R. Haag and J. Roberts\cite{dop} made a general study of identical particle
statistics using the algebraic field theory methods pioneered by Haag.  They found parabose
and parafermi statistics of positive integer orders which as mentioned above were introduced by Green.
They also found another case which they called infinite statistics.  Young patterns label
the inequivalent irreducible representations of the symmetric group.  In parabose (parafermi)
statistics of order $p$ the Young patterns have at most $p$ rows (columns) corresponding to having
at most $p$ particles in an antisymmetric (a symmetric) state.  In infinite statistics all irreducibles
of the symmetric group occur.  Doplicher, et al, did not give an operator realization of infinite
statistics.

\subsection{Infinite statistics}

In 1989 I gave an evening lecture at Wake Forest University.  My talk was attended by
physicists, philosophers, and among people in other disciplines, chemists.  In my talk
I mentioned the bose and fermi commutation relations.  After the talk Roger Hegstrom,
a chemist, asked ``Why not average the bose and fermi commutation relations and consider
the relation
\beee
a(k)a\dggg(l)=\delta(k,l)?"
\eeee
I was surprized to find that such a simple case had not been considered.  Later I found out
that it had, in the mathematical literature, by J. Cuntz\cite{cun}.  With Hegstrom's permission
I developed this case, which turned out to be the first operator example of infinite
statistics\cite{owg-q=0}.  In order to select the Fock-like representation, one must add
the vacuum condition
\beee
a(k)|0\rangle=0.
\eeee
We can calculate all vacuum matrix elements of products of $a$'s and $a\dggg$'s using 
the commutation relation and the vacuum condition.  There is no commutation relation 
involving two $a$'s or two $a\dggg$'s.  There are $n!$ linearly independent $n$-particle
states in Hilbert space if all quantum numbers are distinct; these states differ only
by permutations of the order of the creation operators.  (Later we will see that there are
{\it not} that many independent density matrices or other observables.)  
The matrix of scalar products of these states is the identity matrix,
\beee
M^n_{P,Q}(q)=(Pa\dggg(k_1)a\dggg(k_2) \cdots a\dggg(k_n)|0\rangle,
Qa\dggg(l_1)a\dggg(l_2) \cdots a\dggg(l_n)|0\rangle =\prod_{i=1}^n \delta(k_i,l_i) 
\delta(P,Q)                                             \label{mat}
\eeee
where $P$ and $Q$ are permutations from $S_n$;
that is, the scalar product is zero unless there are the 
same number of creation operators on each side of the scalar product and they have the
same quantum numbers in the same order.  This algebra can be viewed as a deformation of
either the bose or the fermi algebras.  As is typical for deformed algebras, there is
an element that is infinite degree in the generators of the algebra.  In this case
the number operator that obeys
\beee
[n(k), a\dggg(l)]_-=\delta(k,l) a\dggg(l)
\eeee
is the operator of infinite degree; in terms of the $a$'s and the
$a\dggg$'s,
\beee
n(k)=a\dggg(k) a(k) + \sum_t a\dggg(t)a\dggg(k)a(k)a(t) + \sum_{t_1,t_2}
a\dggg(t_2)a\dggg(t_1)a\dggg(k)a(k)a(t_1)a(t_2) + \cdots.
\eeee
There is an analogous formula for the transition operator, $n(k,l)$, that obeys
\beee
[n(k,l),a\dggg(m)]_-=\delta(l,m)a\dggg(k).
\eeee

\section{Quons}

\subsection{The quon algebra}

The quon algebra\cite{ari,pol,owg-q} is the best attempt so far to violate statistics by a small
amount.  The infinite statistics algebra just discussed is the average of the bose and
fermi algebras.  The quon algebra can be obtained as the convex sum of these two algebras,
\beee
\frac{1+q}{2} [a(k), a\dggg(l)]_- + \frac{1-q}{2} [a(k), a\dggg(l)]_+=\delta(k,l),
\eeee
or
\beee
a(k) a\dggg(l) - q a\dggg(l) a(k) =\delta(k,l).                           \label{q}
\eeee
As usual the Fock-like representation is selected by the vacuum condition 
\beee
a(k)|0\rangle =0.                                                          \label{vac}
\eeee
Convexity requires $0 \leq q \leq 1$; for this range the states
have positive squared norms.  Outside this range the squared norms become negative.
Using the algebra (\ref{q}) and the vacuum condition (\ref{vac}) all vacuum matrix elements
of polynomials in the $a$'s and $a\dggg$'s can be calculated; for example,
\[ (a\dggg(k_1)a\dggg(k_2)|0\rangle, a\dggg(l_1)a\dggg(l_2)|0\rangle)=   \]
\[ \delta(k_1,l_1)\delta(k_2,l_2)+q \delta(k_1,l_2)\delta(k_2,l_1)   =   \]
\beee
\frac{1+q}{2}[\delta((k_1,l_1)\delta(k_2,l_2) + \delta(k_1,l_2)\delta(k_2,l_1)]+
\frac{1-q}{2}[\delta((k_1,l_1)\delta(k_2,l_2) - \delta(k_1,l_2)\delta(k_2,l_1)].
\eeee

The first proof of the positivity of the norms was given by D. Zagier\cite{zag},
who gave a tour-de-force calculation of the determinant of the $n! \times n!$
matrix of scalar products (\ref{mat}) for arbitrary $n$,
\beee
det M^n_{P,Q}(q)=\prod_{k=1}^{n-1} (1-q^{k(k+1)})^{\frac{n-k)n!}{k(k+1)}}.
\eeee
As shown above, at $q=0$ the norms are positive and the determinant
is one.  In order for a norm to become negative the determinant has to change
sign.  From Zagier's formula this happens only when $q^{k(k+1)}=1$, i.e., on the
unit circle.  This proves that the norms remain positive between negative one and
one.  

Speicher\cite{spe} gave an ingenious proof of the positivity of the norms using
an ansatz for the Fock-like representation of quons 
analogous to Green's ansatz for parastatistics.  Speicher represented the quon
annihilation operator as the weak operator limit,
\beee
a_k={\rm lim}_{N \rightarrow \infty}N^{-1/2}\sum_{\alpha=1}^N b_k^{(\alpha)},
\eeee
where the $b_k^{(\alpha)}$ are bose oscillators for each $\alpha$, but with
relative commutation relations given by
\beee
b_k^{(\alpha)} b_l^{(\beta) \dagger}=s^{(\alpha, \beta)}b_l^{(\beta) \dagger}
b_k^{(\alpha)}, \alpha \neq \beta,~ {\rm where}~~ s^{(\alpha, \beta)}= \pm 1.
\eeee
This limit is taken as the limit, $N \rightarrow \infty$, in the vacuum expectation state of the 
Fock space representation of the $b_k^{(\alpha)}$.  
In this respect Speicher's ansatz differs from Green's, which is an operator
identity.  To get the Fock-like representation of the quon algebra, 
Speicher chose a probabilistic condition for the signs 
$s^{(\alpha, \beta)}$,
\beee
{\rm prob}(s^{(\alpha, \beta)}=1)=(1+q)/2,
\eeee
\beee
{\rm prob}(s^{(\alpha, \beta)}=-1)=(1-q)/2.
\eeee
Speicher's rules reproduce the quon algebra.  The norms are positive since the sums
of bose or fermi operators have positive norms.  The constraint on $q$ follows because the 
probabilities have to lie between zero and one.

The number and transition operators for general $q$ have infinite degree expansions 
analogous to but more 
complicated than those for the $q=0$ case,
\beee
n(k,l)=a\dggg(k)a(l)+ \frac{1}{1-q^2}\sum_t(a\dggg(t)a\dggg(k)-qa\dggg(k)a\dggg(t))
(a(l)a(t)-qa(t)a(l)) + \cdots.
\eeee
The general formula for the number operator was given by S. Stanciu\cite{sta}.

At $q=\pm 1$ only the symmetric (antisymmetric) representation of ${\cal S}_n$ occurs. 
The quon operators interpolate smoothly between fermi and bose statistics in the
sense that as $q$ departs from $\pm 1$ the vectors formed by polynomials in the creation
operators, which are superpositions of vectors in different irreducible representations
of the symmetric group, have higher weights in the more symmetric (antisymmetric) 
representations, and as $q \rightarrow \mp 1$ the antisymmetric (symmetric) 
representations smoothly become more heavily weighted.  

\subsection{Observables in quon theory}

It is important to note that although there are $n!$ linearly independent
vectors in Fock space associated with a degree $n$ monomial in creation operators that
carry disjoint quantum numbers acting on the vacuum, there
are fewer than $n!$ observables associated with such vectors.  For example, for two
identical quons $1$ and $2$ in orthogonal quantum states the two vectors 
$a\dggg(1)a\dggg(2)|0 \rangle$ and $a\dggg(2)a\dggg(2)|0 \rangle$ are orthogonal and 
each is normalized to one.
Let
\begin{equation}
|\phi_{s,a}\rangle
= N_{s,a} (a^{\dagger}(1) a^{\dagger}(2) \pm a^{\dagger}(2)a^{\dagger}(1))|0\rangle
\end{equation}
be normalized states that are symmetric or antisymmetric under transposition of
$1$ and $2$.
The quon algebra gives
\begin{equation}
N_{s,a}=\frac{1}{\sqrt{2(1 \pm q)}}.
\end{equation}
One can then calculate the expansion
\begin{equation}
a^{\dagger}(1) a^{\dagger}(2)|0\rangle
=\alpha |\phi_{s}\rangle + \beta |\phi_{a}\rangle,
\end{equation}
either using
\begin{equation}
a^{\dagger}(1) a^{\dagger}(2)|0\rangle
=(1/2)[ (a^{\dagger}(1) a^{\dagger}(2) + a^{\dagger}(2) a^{\dagger}(1))
+ (a^{\dagger}(1) a^{\dagger}(2) - a^{\dagger}(2) a^{\dagger}(1))]|0\rangle
\end{equation}
or using
\begin{equation}
a^{\dagger}(1) a^{\dagger}(2)|0\rangle
= \langle \phi_{s}| a^{\dagger}_1 a^{\dagger}_2|0\rangle
|\phi_{s}\rangle
+ \langle \phi_{a}| a^{\dagger}_1 a^{\dagger}_2|0\rangle
|\phi_{a}\rangle.
\end{equation}
Either way gives
\begin{equation}
\alpha = \sqrt{(1+q)/2}, \beta =  \sqrt{(1-q)/2}
\end{equation}
so that 
\beee
a^{\dagger}_1 a^{\dagger}_2|0\rangle=\sqrt{\frac{1+q}{2}}\phi_s + 
\sqrt{\frac{1-q}{2}}\phi_a               \label{sym}
\eeee
and
\beee
a^{\dagger}_2 a^{\dagger}_1|0\rangle=\sqrt{\frac{1+q}{2}}\phi_s -
\sqrt{\frac{1-q}{2}}\phi_a.              \label{ant}
\eeee
Then, dropping the cross terms that are excluded by the superselection
rule separating symmetric and antisymmetric states of identical particles
(and, indeed, states of identical particles in different representations of 
the symmetric group generally),
\begin{equation}
a^{\dagger}_1 a^{\dagger}_2|0\rangle \langle 0 |a_2 a_1 = 
a^{\dagger}_2 a^{\dagger}_1|0\rangle \langle 0 |a_1 a_2
=\frac{1+q}{2}|\phi_{s}\rangle  \langle \phi_{s}|
+\frac{1-q}{2}     |\phi_{a}\rangle \langle \phi_{a}|.       \label{dsym}
\end{equation}
This shows that since particles 1 and 2 are identical
the same observable results
follow when the labels 1 and 2 are transposed.  That also 
shows that the relative phase in Eq.(\ref{sym}) and Eq.(\ref{ant}) 
is not observable.  Equation (\ref{dsym}) states that the
density matrices for $a^{\dagger}_1 a^{\dagger}_2|0\rangle$ and
$a^{\dagger}_2 a^{\dagger}_1|0\rangle$ are {\it identical}, which means that
these two ``states'' correspond to exactly the same physical situation.  We
put quotation marks around the word ``states'' to indicate that these should
really be represented by density matrices.  Note that the sum of 
the coefficients of the two terms in the
two-particle density matrix is one, as it should be.  The general observable
is a linear combination of projectors on the irreducibles of the symmetric
group. 

The parameters $v_F$ and $v_B$ that represent small violations of statistics
can be written in terms of the $q$ parameters; the result is,
\beee
q_F=2v_F-1~{\rm or}~v_F=\frac{1}{2}(1+q_F);~~q_B=1-2v_B~{\rm or}~v_B=
\frac{1}{2}(1-q_B).
\eeee

\subsection{Properties of quon theory}

Surprizingly several properties of relativistic theories that I would expect
to fail for the quon theory (made relativistic kinematically) actually hold.
These include Wick's theorem, cluster decomposition theorems and the $CPT$
theorem.  We are familiar with Wick's theorem for bosons which states that
the vacuum matrix element of a product of free fields is the sum of all possible
products of two-point functions, with each product occuring with factor one.
For fermions the factors are plus or minus one,
depending on the parity of the permutation between the order in the vacuum
matrix element and the order in the product of two point functions.  In 
Wick's theorem for quons the corresponding factors are $q$ raised to the inversion number
of the permutation between the order in the vacuum
matrix element and the order in the product of two point functions.  This result 
reduces to the usual Wick's theorem when $q \rightarrow \pm 1$. The inversion number
can be found conveniently by drawing lines above the vacuum matrix element to
indicate the pairs that are contracted into two-point functions.  The minumum
number of crossings of these lines is the inversion number.  Since both cluster
decomposition and the $CPT$ theorem for vacuum matrix elements of free fields
depend on the properties of two-point functions, these theorems hold for quon
fields.  Note that quon fields, which clearly violate the spin-statistics theorem,
obey the $CPT$ theorem; this emphasises the point made by R. Jost\cite{jos} that
the $CPT$ theorem requires only very weak assumptions. 

If all the usual properties of relativistic field theory hold, then the spin-statistics
theorem holds; thus some property must fail for quons.  The property that does
not hold is locality in the sense of the commutativity of observables at spacelike
separation.  Jost\cite{jos2} showed that if locality holds in an open 
spacelike region then analyticity arguments prove that it holds everywhere outside
the lightcone.  This result does not hold if the violation of
locality decreases--say--exponentially away from the lightcone.  The experimental
bounds on such a violation are not clear.  Note that the nonrelativistic form
of locality
\beee
[\rho({\bf x}), \psi\dggg({\bf y})]_-=\delta({\bf x}-{\bf y}) \psi\dggg({\bf y}),
\eeee
where $\rho$ is the charge density, does hold.

\subsection{Conservation of statistics rules for quon theory}

For the energies of systems that are widely (spacelike) separated to
be additive, all terms in the Hamiltonian must be effective bose operators in
the sense that
\beee
[H(x), \phi(y)]_- \rightarrow 0,~ {\rm as} ~x-y \rightarrow \infty~ {\rm spacelike}
\eeee
for {\it all} fields $\phi$.
This condition imposes ``conservation of statistics'' rules, the simplest of which
is that only an even number of fermi fields can appear in any term of the 
Hamiltonian.  For parafields, which have local observables, Messiah and
I\cite{owgmes} showed that parafields must occur in even degree, except that, 
for $p$ odd, $p$ parafields can occur.   I defined paragrassmann and quongrassmann numbers which
must be used in coupling para and quon operators to external sources and showed that
for external parasources there are analogous
restrictions\cite{owgext}. Using these results, R.C. Hilborn and I\cite{owghil}
gave an heuristic argument relating the $q$ parameter for electrons to that for
photons.  The result,
\beee 
q_e^2=q_{\gamma},
\eeee
allows the very accurate bound from the Ramberg-Snow experiment to be carried
over to photons with comparable accuracy.  Similar arguments work for any particles
that are coupled to electrons through any chain of reactions.

\subsection{Bound states of quons}

The classical result about bound states of bosons and fermion due to 
E.P. Wigner\cite{wig} and to P. Ehrenfest and J.R. Oppenheimer\cite{ehr}
states that a bound state of bosons and fermions is a boson unless it
has an odd number of fermions, in which case it is a fermion.  Hilborn and
I\cite{owghil2} showed that this result generalizes for quons.  A bound
state of $n$ identical quons with parameter $q_{constituent}$ has parameter 
$q_{bound}=q_{constituent}^{n^2}$.  This implies that if $q_{nucleus}$ is
bounded within $\epsilon$ of $\pm 1$ for a nucleus with $A$ nucleons, then
$q_{nucleon}$ is bounded within $\epsilon/A^2$.  Thus the bound on the 
nucleons is {\it stronger} than the bound on the nucleus.  Analogously
the bound on quarks is improved by $1/9$ over the bound on nucleons.
Michael Berry\cite{ber} pointed out that in the context of the quon theory
this result on $q$ for bound states
implies that either the layers of compositeness stop or all particles are
bosons or fermions.

This result reduces to the usual one for bosons and fermions since $q$ and
$q^2$ are even or odd together.

\section{Summary and open questions}

Like any other physical property, statistics should be subjected to 
high-precision experimental tests.  In order to interpret such tests
we need a theory in which statistics can be violated and a parameter
that gives a quantitative measure of the validity of statistics.  A
theory that allows violations of statistics cannot have all the 
properties we might like.  So far quons are the best theory that allows
small violations.

In summary the positive properties of quons as a field theory are
(a) norms are positive, (b) a simple modification of Wick's theorem holds,
(c) cluster decomposition theorems hold, (d) the
$CPT$ theorem holds, and (e) free fields can have relativistic kinematics.
The negative properties are (a) spacelike commutativity
of observables fails, and because of this (b) interacting relativistic
field theory is in doubt.
I do not have a concrete suggestion for the possible origin of small violations 
of the exclusion principle.  One could turn this issue around and observe that 
the constraints of bose
and fermi statistics are grafted onto the general structure of quantum theory in 
an ad hoc way and ask why these constraints are realized in nature.  Study of the 
situation in which these constraints are violated may shed light on why they hold 
for the known particles.  In any case a fundamental issue such as statistics 
should be subjected to experimental tests and to theoretical study, just as 
is being done for Lorentz and $CPT$ invariance.

What we lack is an ``external'' motivation for violation of statistics---that 
is a connection of violations with some other physical property.
We also don't have any insight into the level at which we can expect
violations if they do occur.  Possible external motivations 
for violation of statistics
include (a) violation of $CPT$, (b) violation of locality, (c)
violation of Lorentz invariance, (d) extra space dimensions, (e)
discrete space and/or time and (f)
noncommutative spacetime.  Of these, (a) seems unlikely because
the quon theory which obeys $CPT$ allows violations, (b) seems likely
because if locality is satisfied we can prove the spin-statistics
connection and there will be no violations, (c), (d), (e) and (f) seem
possible.

At the conference the question was raised whether the stability of matter
which depends on the exclusion principle might set stringent bounds
on possible violations of statistics.  This certainly deserves careful
study.  

Hopefully either violations will be found experimentally or
our theoretical efforts will lead to understanding of why only bose
and fermi statistics occur in Nature.


\begin{thebibliography}{99}
\bibitem{pol} Polyakov, A.M., pointed out that the infinite statistics
algebra is a special case of the quantum group relation 
$a(k) \tilde{a}(l)-q \tilde{a}(l)a(k)=\delta(k,l)$, where $\tilde{a}$ is $a \dggg$
for $q$ real and $\tilde{a}_q=a^{\dagger}_{q^{\star}}$ for $q$ complex,
(private communication, 1990) .
\end{thebibliography}
\end{document}